# A statistical theory of complex systems

Jincan Chen[1], Tie Liu[1], Zhifu Huang[2], Guozhen Su[1,*]

[1]Department of Physics, Xiamen University, Xiamen 361005, People Republic of China

[2]College of Information Science and Engineering, Huaqiao University, Xiamen 361021, People Republic of China

**Abstract**. Based on the probability distribution observed in complex systems and an assumption that the probability distributions of complex systems satisfy a new generalized multiplication, it is proved that the statistical theory of complex systems can be established in the analogous extensive framework.

[*]Email. gzsu@xmu.edu.cn

With the development of modern physics, a new probability distribution, $p(x) \propto [1-(1-q)x]^{1/(1-q)} \equiv e_q(-x)$, has been observed in more and more complex physical systems [1-5] such as the driven-dissipative dusty plasma [1,2], non-neutral plasmas [3], dissipative optical lattices [4], test particle transport [5], and atomic momentum [6], where $x$ represents the observable quantity and $q > 0$ is a parameter depending on the performance of a system. This distribution function is different from that of the classical Boltzmann-Gibbs statistics. In classical statistical mechanics, the Boltzmann factor, $exp(-x)$, is obtained under some ideal conditions. Obviously, the theory of the Boltzmann-Gibbs statistics cannot be successfully used to interpret some observed results of complex systems with long-range interaction [6,7], long-duration memory [8,9], fractal phase space structures [10], and nonsmooth space-time structures such as super string theory and the corresponding informational entropy [11-13]. In addition, several existing nonextensive statistical theories [14, 15] cannot satisfactorily explain the observed results of complex systems. Thus, it urgently needs to develop some new statistic theories.

It is found without difficulty that the probability distribution observed in complex systems can be expressed as $p_i = \frac{1}{Z}[1-(J-1)x_i]^{1/(J-1)}$, where $J = 2-q < 2$ is a parameter describing the performance of complex systems and the partition function $Z = \sum_i [1-(J-1)x_i]^{1/(J-1)}$. The value of $Z$ depends on the random variable distribution $\{x_i\}$ and $J$. If a spectrum shift ($x_i \to x_i - a$) is adopted so that $Z = \sum_i [1-(J-1)(x_i-a)]^{1/(J-1)} = 1$, the probability distribution mentioned above can be rewritten as

$$p_i = [1-(J-1)(x_i-a)]^{1/(J-1)}. \tag{1}$$

It is interesting to note that a variational relationship (VarEntropy method) between the random variable and the uncertainty measure of a system, $ds = d\langle x \rangle - \langle dx \rangle = \sum_i x_i dp_i$, was proposed [16] to discuss the probability distribution function of a system. Because the "VarEntropy" is exactly the reverse process of "MaxEnt"[17], the uncertainty measure can be used to derive the entropy of a system [18]. In principle, one can get different entropic forms for various distribution functions by using the VarEntropy method. Using Eq. (1) and the variational relationship mentioned above, one can derive the entropy expression of complex systems as

$$S = (1 - \sum_i p_i^J)/[J(J-1)], \ (0 < J < 2), \tag{2}$$

where the constraint condition $J > 0$ ensures $S > 0$.

Using Eq. (2), the normalized condition $\sum_i p_i = 1$, the expectation of an observable quantity $\langle O \rangle = \sum_i p_i O_i$, and the Lagrange multiplier method, one can derive the probability distribution function of complex systems as

$$p_i = [1 - (J-1)\beta(O_i - A)]^{1/(J-1)}, \tag{3}$$

where $A$ is a parameter determined by the normalized condition $\sum_i [1 - (J-1)\beta(O_i - A)]^{1/(J-1)} = 1$, $\beta$ is also a parameter determined by the property of the system, and $O_i$ represents a possible value of an observable quantity in complex systems. As long as $\beta O_i = x_i$ and $\beta A = a$ are chosen, Eq. (3) is exactly identical with Eq. (1). It shows clearly that Eq. (2) can be directly used to describe the statistical properties of complex systems.

It is worthwhile to point out that Eq. (2) is formally similar to the Tsallis entropy [14],

$$S^T = (1 - \sum_i p_i^q)/(q-1), \ (q \in R) \tag{4}$$

but they are essentially different from each other. It is well known that the Tsallis entropy and three expressions of the expectations of observable quantities [19], (i) $\langle O \rangle = \sum_i p_i O_i$, (ii) $\langle O \rangle = \sum_i p_i^q O_i$, and (iii) $\langle O \rangle = \sum_i p_i^q O_i / (\sum_j p_j^q)$ are some foundations of the Tsallis statistical mechanics. It has been found that the second definition of the expectations of observable quantities is in conflict with the expectation of a constant $C_1$, i.e., $C_1 \neq \sum_i p_i^q C_1$ $(q \neq 1)$. Thus, the second definition has been seldom adopted. It has been also proved that the third definition of the expectations is neither stable [20, 21] nor in consistent with the generalized Stosszahlansatz and the associated $H$-theorem [22]. It means that the third definition of the expectation is not experimental robustness or observable so that the corresponding results cannot be compared with the experimental observations. By using the Lagrange multiplier method, the Tsallis entropy, and the first definition of the expectations, the probability distribution $p_i = \left\{ \sum_j p_j^q - \frac{q-1}{q} \beta (O_i - \langle O \rangle) \right\}^{\frac{1}{q-1}}$ can be derived [23]. However, the probability distribution is different from that observed in complex physical systems. Although the probability distribution may be expressed as $p_i = (1/Z^T)[1 - (q-1)x_i]^{1/(q-1)}$, which is similar to the form of the probability distribution observed in complex systems, the $x_i$ in this expression are dependent on the parameter $q$ and not a possible value of an observable quantity, because $x_i = \beta O_i / [\sum_j q p_j^q + (q-1)\beta \langle O \rangle]$ and $Z^T = 1/(\sum_j p_j^q + \frac{q-1}{q} \beta \langle O \rangle)^{1/(q-1)}$. It shows clearly that Eq. (2) can be used to better describe the performance of a complex system than Eq. (4). On the other hand, it is very worthy to note that the Tsallis statistical mechanics was developed on the basic of the assumption [14] that the probability distribution $p_{ij}(C)$ of a coupling system $C$ consisting of two independent subsystems $A$ and $B$ is equal to the production of the probability distributions $\{p_i(A)|i=1,2,...W_A\}$ and $\{p_j(B)|j=1,2,...W_B\}$ of the two subsystems, i.e.,

$$p_{ij}(C) = p_i(A)p_j(B), \qquad (5)$$

where $W$ is the possible number of the states. In general, Eq. (5) cannot be directly used to describe the probability distribution relation between the coupling system and the two complex subsystems [24-26]. The analyses above show that the Tsallis nonextensive statistical theory is facing with the dual challenges of experimental observations and basic theories.

Some researchers suggested that for a coupling system composed of two complex subsystems $A$ and $B$, the probability distribution should be given by [24, 25]

$$p_{ij}(C) = p_i(A)p_{ij}(B|A) + p_{ij}(A|B)p_j(B), \qquad (6)$$

where $p_{ij}(A|B)$ and $p_{ij}(B|A)$ are the conditional probabilities. Equation (6) seems to be more reasonable, but it is difficult in giving the concrete form of $p_{ij}(A|B)$ and $p_{ij}(B|A)$ except a limited class of systems [27, 28], so that the thermostatistic properties of many complex systems are hardly investigated. Thus, in the process to develop the statistical theory of complex systems, one primary task is to find a general relation between the probability distributions of the coupling system and two subsystems.

According to the distribution function of the classical Boltzmann-Gibbs statistics, the entropy of a coupling system $C$ consisting of two independent subsystems $A$ and $B$ is extensive. However, it is seen from Eq. (2) that the entropies of complex systems are dependent on the parameter $J$. Thus, it is assumed that based on Eq. (2) derived from the observed results of complex systems, the entropy $S(C)$ of a coupling system $C$ consisting of two complex subsystems $A$ and $B$ may be expressed as

$$J(J-1)S(C) = J_1(J_1-1)S(A) + J_2(J_2-1)S(B), \tag{7}$$

which implies

$$1 - \sum_{ij} p_{ij}^J(C) = 2 - \sum_i p_i^{J_1}(A) - \sum_i p_j^{J_2}(B), \tag{7a}$$

where $S(A)$ and $S(B)$ are the entropies of two subsystems, $J_1$ and $J_2$ are two parameters describing the performance of complex subsystems, and $J$ is a parameter describing the performance of the coupling system. When $J_1 = 1$ or $J_2 = 1$, the subsystem A or B becomes a simple system and Eq. (7) is simplified as

$$S(C) = \frac{1}{J(J-1)}(1 - \sum_{ij} p_{ij}^J) = \begin{cases} \frac{J_2(J_2-1)}{J(J-1)} S(B) = \frac{1}{J(J-1)}(1 - \sum_i p_j^{J_2}), & (J_1 = 1) \\ \frac{J_1(J_1-1)}{J(J-1)} S(A) = \frac{1}{J(J-1)}(1 - \sum_i p_i^{J_1}), & (J_2 = 1) \end{cases}. \tag{8}$$

In such a case, the entropy of the subsystem A or B is taken as a datum point of the entropy of the coupling system. When $J_1 = J_2 = 1$, it necessarily has $J = 1$ and complex systems reduce to simple systems, whose entropies are extensive.

It can be found from Eq. (7a) that the relation between the probability distributions of the coupling system and two subsystems is given by

$$\sum_{ij} p_{ij}^J = \sum_{ij} (p_i^{J_1} p_j + p_i p_j^{J_2} - p_i p_j). \tag{9}$$

Using Eqs. (3) and (9), one can further derive the relations between the expectations of observable quantities of the coupling system and two subsystems as

$$\langle O_{ij}(C)\rangle = \sum_{ij} p_{ij} O_{ij} = \sum_{ij} p_{ij}[\frac{1}{(J-1)\beta}(1-p_{ij}^{J-1})+C]$$

$$= \frac{(J_1-1)\beta_1}{(J-1)\beta}\sum_i p_i[\frac{1}{(J_1-1)\beta_1}(1-p_i^{J-1})+A] + \frac{(J_2-1)\beta_2}{(J-1)\beta}\sum_j p_j[\frac{1}{(J_2-1)\beta_2}(1-p_j^{J-1})+B]\}, \quad (10)$$

$$= \frac{J_1-1}{J-1}\frac{\beta_1}{\beta}\sum_i p_i O_i + \frac{J_2-1}{J-1}\frac{\beta_2}{\beta}\sum_j p_j O_j = \frac{J_1-1}{J-1}\frac{\beta_1}{\beta}\langle O_i(A)\rangle + \frac{J_2-1}{J-1}\frac{\beta_2}{\beta}\langle O_j(B)\rangle$$

where $C = \frac{J_1-1}{J-1}\frac{\beta_1}{\beta}A + \frac{J_1-1}{J-1}\frac{\beta_2}{\beta}B$. When $\beta_1 = \beta_2 = \beta$, Eq. (10) is simplified as

$$\langle O_{ij}(C)\rangle = \frac{J_1-1}{J-1}\langle O_i(A)\rangle + \frac{J_2-1}{J-1}\langle O_j(B)\rangle. \tag{11}$$

Similarly, when $J_1 = 1$ or $J_2 = 1$, the expectations of observable quantities of the subsystem A or B are taken as a datum point of the expectations of observable quantities of the coupling system. When $J_1 = J_2 = 1$, the expectations of observable quantities of systems are extensive.

The results obtained above show that if the entropy of a coupling system consisting of two complex subsystems is given by Eq. (7), which is analogous extensive, a general relation, Eq. (9), between the probability distributions of the coupling system and two subsystems can be uniquely determined and it can be easily proved that the expectations of observable quantities are also analogous extensive. On the other hand, if Eq. (9) is taken as a new assumption to replace Eq. (5), Eqs. (7) and (10) can be directly derived from Eq. (2). It shows that the entropies and expectations of observable quantities of complex systems can be calculated in the analogous extensive framework.

Obviously, Eq. (9) is relative complex so that the probability distribution $\{p_{ij}\}$ of the coupling system cannot be conveniently obtained from the known probability distributions $\{p_i\}$ and $\{p_j\}$ of two subsystems. However, based on Eq. (9), it may be assumed that the probability distribution functions of complex systems satisfy a new generalized multiplication, i.e.,

$$p_{ij}(J) = p_i(J_1) \otimes p_j(J_2) = (p_i^{J_1} p_j + p_i p_j^{J_2} - p_i p_j)^{1/J}, \tag{12}$$

where $\otimes$ indicates a generalized multiplication and the concrete value of $J$ is determined by the normalized condition. Equation (12) is one rational generalization of Eq. (9) and can be conveniently used to calculate the probability distributions $\{p_{ij}\}$ of the coupling system with different parameters $J_1$ and $J_2$. Using Eqs. (2) and (12), one can easily derive Eqs. (7) and (10) without any other additional condition. It is worthwhile to point out that when $J_1 = J_2 \neq 1$, the concrete value of $J$ still needs to be determined by the normalized condition. It means that if Eq. (12) is taken as the starting point, Eqs. (7) and (10) may be used to discuss the statistical properties of a coupling system consisting of two complex subsystems.

The results obtained above can be summarized as follows. Based on the probability distribution observed in complex systems, a generic entropy expression of complex systems with a parameter $J$ depending on the systemic performance is derived without any additional condition. A probability distribution is derived by using the Lagrange multiplier method and then proved to be exactly identical with the probability distribution observed in complex systems. With the help of the assumption that the entropies of complex systems are analogous extensive, the probability distribution of the coupling system consisting of two complex systems with different parameters $J_1$ and $J_2$ is derived and it is proved that the expectations of observable quantities of complex systems are also analogous extensive. Finally, it is proved that as long as a new generalized multiplication of probability distribution functions is introduced, the statistical theory of complex systems can be established in the analogous extensive framework.

**Acknowledgments**

This work has been supported by the National Natural Science Foundation (No. 11175148), People's Republic of China.